\documentclass{article}

     \usepackage[final, nonatbib]{neuripsDeepInverse_2021}

\usepackage[utf8]{inputenc} 
\usepackage[T1]{fontenc}    
\usepackage{hyperref}       
\usepackage{url}            
\usepackage{booktabs}       
\usepackage{amsfonts}       
\usepackage{nicefrac}       
\usepackage{microtype}      
\usepackage{xcolor}         

\usepackage{subcaption}
\usepackage{caption}
\usepackage{subcaption}
\usepackage{amsmath}
\usepackage{graphicx}
\usepackage{wrapfig}
\usepackage[nottoc]{tocbibind}

\title{Sparse deep computer-generated\\ holography for optical microscopy}

\author{Alex Liu \qquad \quad Yi Xue* \qquad \quad Laura Waller\\
  Department of Electrical Engineering and Computer Sciences\\
  University of California, Berkeley\\
  *corresponding author: \texttt{xueyi@berkeley.edu}
}

\begin{document}

\maketitle

\begin{abstract}
  Computer-generated holography (CGH) has broad applications such as direct-view display, virtual and augmented reality, as well as optical microscopy. CGH usually utilizes a spatial light modulator that displays a computer-generated phase mask, modulating the phase of coherent light in order to generate customized patterns. The algorithm that computes the phase mask is the core of CGH and is usually tailored to meet different applications. CGH for optical microscopy usually requires 3D accessibility (i.e., generating overlapping patterns along the $z$-axis) and micron-scale spatial precision. Here, we propose a CGH algorithm using an unsupervised generative model designed for optical microscopy to synthesize 3D selected illumination. The algorithm, named sparse deep CGH, is able to generate sparsely distributed points in a large 3D volume with higher contrast than conventional CGH algorithms. 
\end{abstract}

\section{Introduction}

Computer-generated holography (CGH) has become a powerful and versatile technique for 3D display, particularly recently with advances in computation. CGH is able to generate a desired 3D illumination pattern by modulating the phase of light in the pupil plane (Fourier domain), typically with a liquid crystal-based spatial light modulator (SLM). The quality of the 3D illumination pattern depends on the computer-generated phase mask displayed on the SLM. Synthesizing 3D illumination patterns in the real domain with only 2D phase control in the Fourier domain is an ill-posed problem that is usually solved by iterative optimization algorithms~\cite{Fienup1978-so,Ripoll2004-jw}. For example, the Gerchberg-Saxton (GS) algorithm \cite{Fienup1978-so}, is a classic phase retrieval method that iteratively optimizes the phase in the Fourier domain for the target intensity in the real domain. The GS algorithm is straightforward and robust, but it usually generates low contrast images and slow computational speed due to iterations. The computed phase mask is also prone to get stuck in local minimum because this is a non-convex problem.

Recently, deep learning based CGH algorithms have been shown to improve computational speed with comparable or even higher contrast than classic algorithms~\cite{Peng2020-xq, Shi2021-zd, Hossein_Eybposh2020-lz, Horisaki2018-le}. Most of the deep learning CGH algorithms aim for near-eye display and augmented/virtual reality; that is, the targeted illumination patterns are 2D or pseudo 3D (i.e., 2D plus depth, such that overlap in the axial $z$-axis is not allowed)~\cite{Peng2020-xq, Shi2021-zd}. However, microscope CGH algorithms used for holographic photo-stimulation \cite{Chen2018-nl} and imaging of neurons \cite{Yang2018-zm} requires the ability to address neurons that are distributed in 3D with possible $z$ overlap. Further, neural photo-stimulation applications require micro-scale spatial accuracy, which is much higher resolution than CGH for display purpose.  This makes CGH design for neural photo-stimulation more challenging; however, these methods also have the advantage of the target illumination being a sparse point cloud in 3D, and neuroscientists can flexibly adjust the sparsity of fluorescence labeling of neurons. 
\begin{wrapfigure}[15]{r}{0.5\textwidth}
  \begin{center}
  \includegraphics[width=0.5\textwidth]{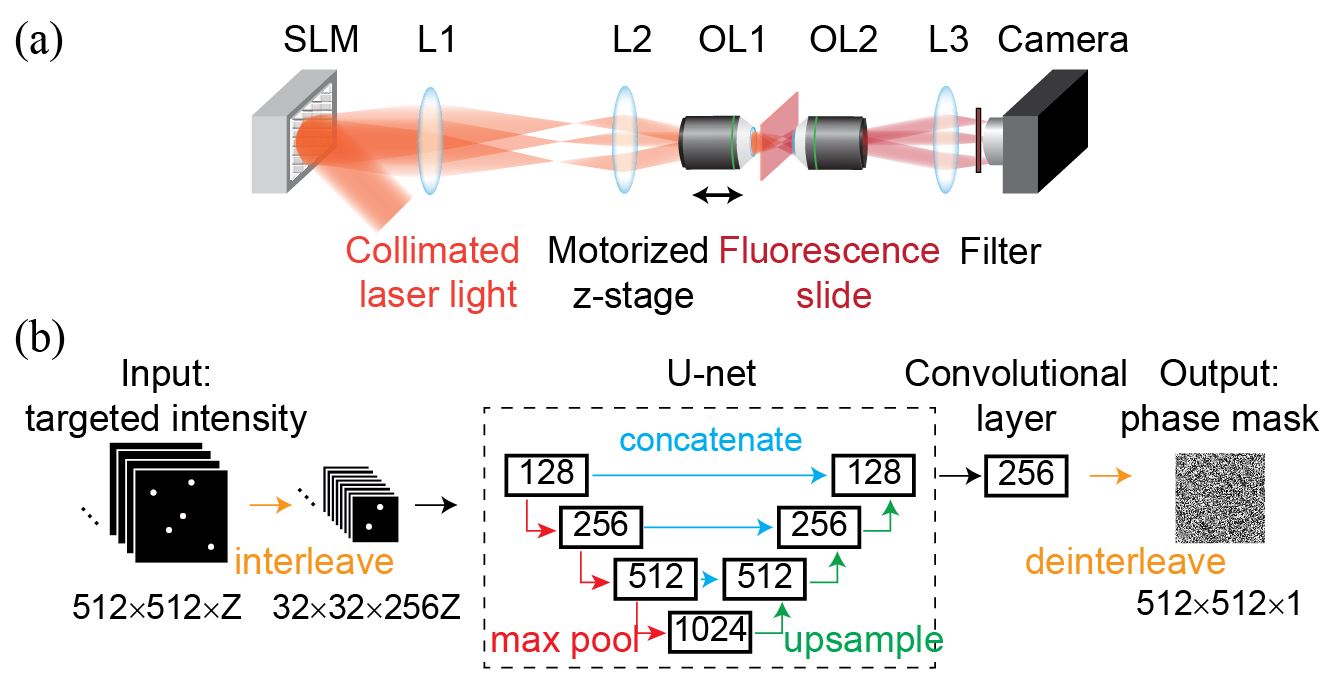}
  \end{center}
  \caption{(a) Schematic diagram of the experimental setup. SLM, spatial light modulator; L, lens. (b) The model architecture. The input is the desired 3D intensity and the output is the corresponding 2D phase mask to be displayed on the SLM.}
  \label{fig0}
\end{wrapfigure}
Given these requirements, we designed a deep learning based CGH specifically for sparse 3D photo-stimulation for neurons. The algorithm utilizes an unsupervised generative approach using a U-net with customized loss function to increase the image contrast, named Sparse Deep CGH (SDCGH). We experimentally show that 3D illumination patterns generated by our algorithm have higher contrast than that generated by classic algorithms.

\section{Model Specification}
\subsection{Wave Propagation Model}
A collimated laser beam, $A_{Laser}(x,y)$, is modulated by a discretized 2D phase mask $\phi_{SLM}$ on the SLM (see Fig.1a). The output complex-field after the modulation can be expressed as $P_{SLM} = A_{Laser}(x, y)e^{i\phi_{SLM}(x,y)}$. After propagating through the relay lenses (L1, L2), the field is demagnified by a factor of M ($M=f_2/f_1$, $f_1$ is the focal length of L1 and $f_2$ is the focal length of L2) to match the size of the pupil of the objective lens, labeled as $P_{SLM}'=MP_{SLM}$.  The complex-field at the image plane, $z=0$, is determined by a 2D Fourier Transform of the complex field $P_{SLM}'$ at the Fourier plane and is expressed using the discretized Fraunhofer wave propagation model \cite{goodman2005}:
\begin{equation}
    P(x,y,z=0) = \frac{1}{i\lambda f}\iint P_{SLM}'(x',y')\text{exp}\left[\frac{-2i\pi(xx' + yy')}{\lambda f}\right]dx'dy',
\end{equation}
where $\lambda$ is wavelength and $f$ is the focal length of the objective lens OL1. The phase and amplitude of the complex field at the image plane $z=0$ determine each subsequent discretized image plane across the axial $z$ dimension by the Fresnel wave propagation equation \cite{goodman2005}:
\begin{equation}
    P(x,y,z) = \frac{e^{ikz}}{i\lambda z}\iint{P(x',y', 0)}\text{exp}\left[\frac{i\pi((x-x')^{2} + (y-y')^{2})}{\lambda z}\right]dx'dy',
\end{equation}
where $k=2\pi /\lambda$. We represent this wave propagation model based on the SLM phase mask $\phi_{SLM}$ using a differentiable transfer function from the SLM plane to the image plane such that it can be used with the automatic differentiation engine in PyTorch.

\subsection{Generative Model}
Our generative model is based on the U-Net network architecture \cite{ronneberger2015} and a pair of interleaving/de-interleaving layers \cite{xiao2018, Hossein_Eybposh2020-lz} (Fig. 1b). The model takes in as input a 512 $\times$ 512 $\times$ Z target intensity pattern and outputs a corresponding 512 $\times$ 512 phase mask, where $Z$ is the number of image planes in the $z$-axis. We utilize an accuracy metric $\alpha$ based on the $L_{2}$ norm for our loss function, defined as
\begin{equation}
    \alpha(I_{\phi}, V) = \frac{\Sigma_{x,y,z}I_{\phi}(x,y,z)V(x,y,z)}{\sqrt{[\Sigma_{x,y,z}V(x,y,z)^2][\Sigma_{x,y,z}I_{\phi}(x,y,z)^2]}},
\end{equation}

\noindent where $I_{\phi}$ is the simulated 3D illumination pattern based on the wave propagation model using the generated phase mask $\phi$ and $V$ is the target 3D illumination pattern. The loss, interpreted as the difference in intensity distribution between the simulated and target 3D illumination patterns, is propagated through the wave propagation model and subsequently passed to the neural network model parameters to facilitate learning. The accuracy is differentiable with respect to the model parameters and is implemented using automatic differentiation for real and complex tensors in PyTorch.

The volume-preserving interleaving and de-interleaving layers prepend and append the U-net style architecture, respectively. Time complexity of convolutional networks is linearly proportional to the spatial resolution of input features \cite{xiao2018, chaitanya2017, nalbach2016}. The interleaving layer reduces time complexity and preserves the total voxel information by decreasing the spatial resolution, while correspondingly increasing the number of channels \cite{xiao2018}. The interleaving layer transforms an input with dimension $(H,W,C)$ to an output with dimension $(\frac{H}{r},\frac{W}{r},C\times r^{2})$, where $r$ is the interleaving factor. The de-interleaving layer is the transpose of the interleaving layer, which reshapes the $(\frac{H}{r},\frac{W}{r},C\times r^{2})$ input to the $(H, W, C)$ output. In our experiments, we used an interleaving factor of $r = 16$ to best tradeoff between computational time and reconstruction quality. The input to the interleaving layer is the $(512,512,Z)$ 3D target illumination pattern and the layer returns a $(32, 32, Z\times16^{2})$ vector to be input to the U-net style architecture. The output of the U-net style architecture is a vector with dimension $(32\times32\times256)$, note that the dimension is invariant to $Z$, which is rearranged through the de-interleaving layer to return a $512\times512$ phase mask.

The neural network contained within the interleaving and de-interleaving layers follows a U-net style architecture with an additional convolutional layer to upsample the dimension. The U-net architecture contains three encoder blocks, three decoder blocks, and a bottleneck block in between. Each of the encoder and bottleneck blocks increase the channel dimension by a factor of $2\times$, while the decoder blocks decrease the channel dimension by a factor of $2\times$. The encoder and bottleneck blocks consist of a sequence of convolutional layers that increase the channel dimension by a factor of $2\times$, a batch-normalization layer, a ReLU activation layer, then another convolutional layer that preserves the channel dimension, a batch-normalization layer, and ReLU activation layer. The decoder blocks also follow this sequence with the only difference in that the first convolutional layer in the sequence decreases the channel dimension by a factor of two. Each encoder block is directly followed by a max-pooling layer that decreases the spatial dimension of the output tensor by a factor of $2\times$. The bottleneck and each decoder block except the last is directly followed by an upsampling layer that increases the spatial dimension of the output tensor by a factor of $2\times$. The output of each encoder block is concatenated along the channel dimension to the input of the corresponding decoder block of the same size. The last encoder block is followed by an additional convolutional layer with kernel size $(1, 1)$ that doubles the channel dimension to maintain consistent total voxel information with the 512 $\times$ 512 output phase mask as a result of the de-interleaving layer.

The real-valued phase mask output $\phi_{SLM}(x,y)$ from the generative model is then used along with the static amplitude $A_{Laser}$ specified in simulation to define the modulated complex wave $P_{SLM}(x,y) = A_{Laser}(x, y)e^{i\phi_{SLM}(x,y)}$ that propagates through the simulated optical system determined by Eqs. (1) and (2). We implement this with a differentiable transfer function from the SLM plane to the image plane so that the loss from the difference between the simulated and target 3D illumination pattern can be used for gradient-based optimization of the generative model parameters.

\section{Experimental Results}
\label{headings}
We build an optical setup to experimentally demonstrate SDCGH and compare it with GS. As shown in Fig. 1b, a SLM ($1920\times1080$ pixels) displaying a phase mask modulates collimated laser light at 589nm wavelength. After passing through the relay lenses (L1 and L2, f1=150mm, f2=180mm), light is focused by the objective lens (10x, NA 0.45) to generate a 3D point cloud. To image the 3D point cloud, we mount the excitation objective lens on a motorized stage to move it along the $z$-axis. We use a thin fluorescence slide combined with sCMOS camera to capture fluorescence light that is excit ed by the point cloud at each depth. The fluorescence images of the 3D point cloud generated by GS and SDCGH, respectively, are shown in Fig. 2. 

\begin{figure}[h]
  \includegraphics[width=\linewidth]{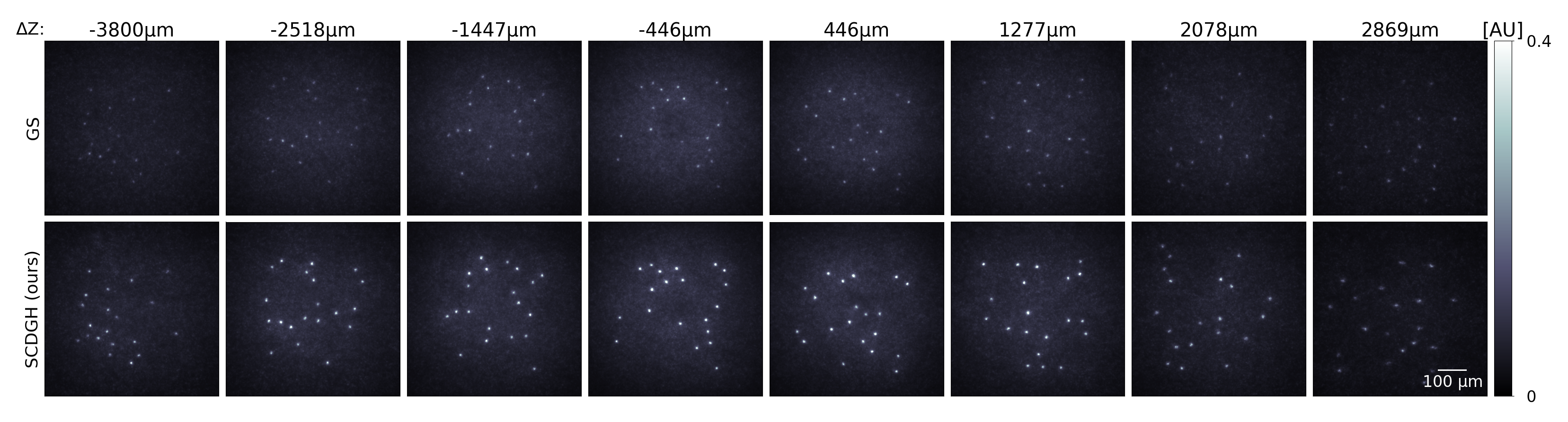}
  \caption{Comparison of fluorescence images generated by GS and SDCGH.Each column shows the targets at each axial plane. The targets generated by SDCGH have higher contrast than that generated by GS.}
  \label{fig2}
\end{figure}

Our model generates accurate phase masks by first training on the desired 3D illumination pattern. For a target intensity pattern with 20 illuminated points sparsely distributed across 8 $z$-planes, we see that the SDCGH-based reconstructed 3D image has overall higher intensity in the target point areas than the GS-based reconstruction, while maintaining a similar low intensity distribution in the background non-target areas (Fig. 2, Fig. 3a). This is also evident by the comparison of the peak signal-to-noise ratio (PSNR) between the images generated by the two methods. The reconstructed 3D illumination pattern generated by SDCGH has PSNR approximately 5 dB higher than that of GS over all 8 $z$-planes (Fig. 3b), indicating that SDCGH offers higher quality in generation of illumination patterns. SDCGH also achieves high spatial resolution especially in the radial direction, as shown in Fig. 3c-d. Besides sparse targets, we also generated dense targets (100 illuminated points per plane) using SDCGH and GS, respectively. Compared to the dense targets generated by GS, the dense targets generated by SDCGH have similar contrast (Fig. 3e). Therefore, our algorithm works best for sparse distributed targets in 3D. 

\begin{figure}[h]
  \includegraphics[width=\linewidth]{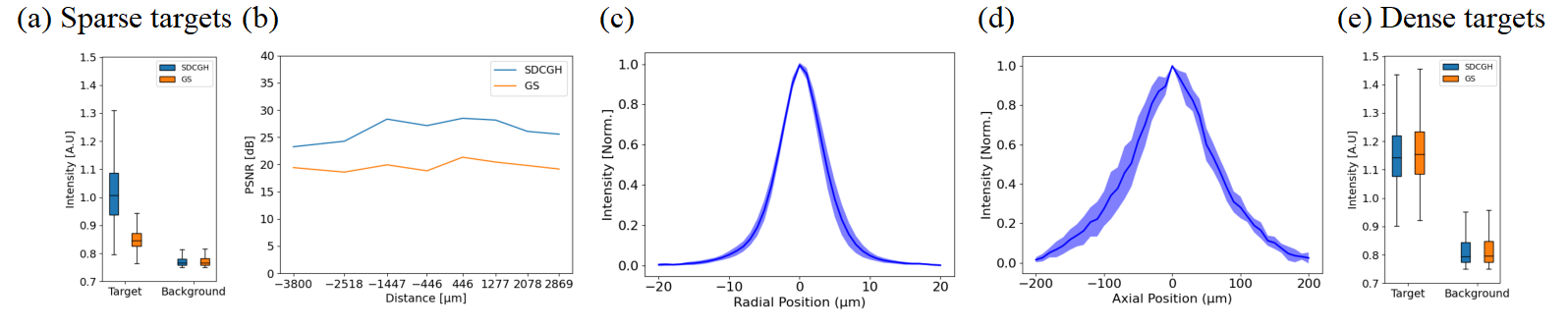}
  \caption{(a) Intensity distribution comparison between SDCGH and GS generated sparse illumination patterns, respectively. The sparse targets generated by SDCGH have higher intensity over that of GS, while the background intensity is roughly the same. (b) Per $z$-plane PSNR comparison between SDCGH and GS, (c) radial and (d) axial point spread function (PSF) of target points. (e) Intensity distribution comparison between SDCGH and GS generated dense illumination patterns, respectively. The dense targets generated by both methods have similar intensity. }
  \label{fig2}
\end{figure}

Our model was also trained and evaluated on 3D illumination patterns with target points overlapping in $z$. For a pattern with 20 illuminated points per $z$-plane, five of which are overlapping, the constructed 3D intensity pattern from the SDCGH-generated phase mask displays all target illumination points (Fig. 4a). Additionally, the displayed illumination pattern maintains low intensity for intermediate $z$-planes not specified by the discretized 3D target pattern. More precisely, the $(x, y)$ positions of overlapping target points have high intensity for specified $z$-planes, but also has desired low intensity in the same $(x, y)$ positions at intermediate $z$-slices (Fig. 4b).

\begin{figure}[h]
  \includegraphics[width=\linewidth]{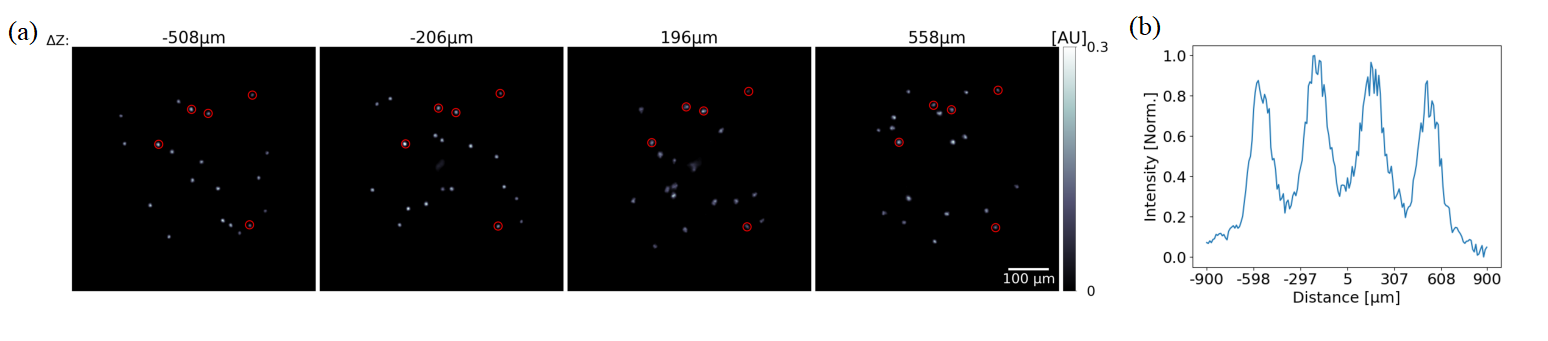}
  \caption{(a) Fluorescence images of SDCGH-generated illumination pattern with overlapping points in $z$ circled. (b) Intensity distribution of overlapping points across the $z$-axis.}
  \label{fig2}
\end{figure}

\section{Conclusion}

In this work we demonstrated and validated SDCGH, a deep learning based algorithm for computer generated holography that is specifically tailored to generate sparse point clouds in 3D for large-scale holographic microscopy. Our approach is demonstrated for generating holographic patterns across 8 $z$-planes allowing superposition, and achieves much higher contrast than conventional methods. This algorithm will improve CGH performance for large-scale holographic microscopy in neuronal imaging and photo-stimulation of the brain. 

\bibliographystyle{unsrt}
\bibliography{SDCGH_neurips_2021}

\begin{thebibliography}{10}

\bibitem{Fienup1978-so}
J~R Fienup.
\newblock Reconstruction of an object from the modulus of its fourier
  transform.
\newblock {\em Opt. Lett., OL}, 3(1):27--29, July 1978.

\bibitem{Ripoll2004-jw}
Olivier Ripoll, Ville Kettunen, and Hans~Peter Herzig.
\newblock Review of iterative fourier-transform algorithms for beam shaping
  applications.
\newblock {\em Organ. Ethic.}, 43(11):2549--2548, November 2004.

\bibitem{Peng2020-xq}
Yifan Peng, Suyeon Choi, Nitish Padmanaban, Jonghyun Kim, and Gordon Wetzstein.
\newblock Neural holography.
\newblock In {\em {ACM} {SIGGRAPH} 2020 Emerging Technologies}, number Article
  8 in SIGGRAPH '20, pages 1--2, New York, NY, USA, August 2020. Association
  for Computing Machinery.

\bibitem{Shi2021-zd}
Liang Shi, Beichen Li, Changil Kim, Petr Kellnhofer, and Wojciech Matusik.
\newblock Towards real-time photorealistic {3D} holography with deep neural
  networks.
\newblock {\em Nature}, 591(7849):234--239, March 2021.

\bibitem{Hossein_Eybposh2020-lz}
M~Hossein~Eybposh, Nicholas~W Caira, Mathew Atisa, Praneeth Chakravarthula, and
  Nicolas~C P{\'e}gard.
\newblock {DeepCGH}: {3D} computer-generated holography using deep learning.
\newblock {\em Opt. Express}, 28(18):26636--26650, August 2020.

\bibitem{Horisaki2018-le}
Ryoichi Horisaki, Ryosuke Takagi, and Jun Tanida.
\newblock Deep-learning-generated holography.
\newblock {\em Appl. Opt.}, 57(14):3859--3863, May 2018.

\bibitem{Chen2018-nl}
I-Wen Chen, Eirini Papagiakoumou, and Valentina Emiliani.
\newblock Towards circuit optogenetics.
\newblock {\em Curr. Opin. Neurobiol.}, 50:179--189, June 2018.

\bibitem{Yang2018-zm}
Weijian Yang and Rafael Yuste.
\newblock Holographic imaging and photostimulation of neural activity.
\newblock {\em Curr. Opin. Neurobiol.}, 50:211--221, June 2018.

\bibitem{goodman2005}
Joseph~W Goodman.
\newblock Introduction to fourier optics.
\newblock {\em Introduction to Fourier optics, 3rd ed., by JW Goodman.
  Englewood, CO: Roberts \& Co. Publishers, 2005}, 1, 2005.

\bibitem{ronneberger2015}
Olaf Ronneberger, Philipp Fischer, and Thomas Brox.
\newblock U-net: Convolutional networks for biomedical image segmentation.
\newblock {\em CoRR}, abs/1505.04597, 2015.

\bibitem{xiao2018}
Lei Xiao, Anton Kaplanyan, Alexander Fix, Matthew Chapman, and Douglas Lanman.
\newblock Deepfocus: Learned image synthesis for computational displays.
\newblock {\em ACM Trans. Graph.}, 37(6), December 2018.

\bibitem{chaitanya2017}
Chakravarty R.~Alla Chaitanya, Anton~S. Kaplanyan, Christoph Schied, Marco
  Salvi, Aaron Lefohn, Derek Nowrouzezahrai, and Timo Aila.
\newblock Interactive reconstruction of monte carlo image sequences using a
  recurrent denoising autoencoder.
\newblock {\em ACM Trans. Graph.}, 36(4), July 2017.

\bibitem{nalbach2016}
Oliver Nalbach, Elena Arabadzhiyska, Dushyant Mehta, Hans{-}Peter Seidel, and
  Tobias Ritschel.
\newblock Deep shading: Convolutional neural networks for screen-space shading.
\newblock {\em CoRR}, abs/1603.06078, 2016.

\end{thebibliography}

\end{document}